\PassOptionsToPackage{pdfpagelabels=false}{hyperref} 
\documentclass[a4paper, 10pt, conference]{IEEEtran}
\IEEEoverridecommandlockouts
\IEEEpubid{\makebox[\columnwidth]{978-1-7281-5919-5\/20\/\$31.00~\copyright~2020 IEEE \hfill} \hspace{\columnsep}\makebox[\columnwidth]{ }}
\IEEEpubidadjcol

\begin{document}

\title{Personalised Visual Art Recommendation by Learning Latent Semantic Representations}

\author{
\IEEEauthorblockN{Bereket Abera Yilma\IEEEauthorrefmark{1}\IEEEauthorrefmark{2},  Najib Aghenda\IEEEauthorrefmark{1}, Marcelo Romero\IEEEauthorrefmark{1}\IEEEauthorrefmark{2} Yannick Naudet\IEEEauthorrefmark{1} and Herv\'{e} Panetto\IEEEauthorrefmark{2}}
\IEEEauthorblockA{\IEEEauthorrefmark{1}Luxembourg Institute of Science and Technology (LIST)}
\IEEEauthorblockA{\IEEEauthorrefmark{2}Université de Lorraine, CNRS, CRAN}
\IEEEauthorblockA{\{bereket.yilma, Najib.aghenda, marcelo.romero, yannick.naudet\}@list.lu, herve.panetto@univ-lorraine.fr}
}

\maketitle

\begin{abstract}

In Recommender systems, data representation techniques play a great role as they have the power to entangle, hide and reveal explanatory factors embedded within datasets. Hence, they influence the quality of recommendations. Specifically, in Visual Art (VA) recommendations the complexity of the concepts embodied within paintings, makes the task of capturing semantics by machines far from trivial. In VA recommendation, prominent works commonly use manually curated metadata to drive recommendations. Recent works in this domain aim at leveraging visual features extracted using Deep Neural Networks (DNN). However, such data representation approaches are resource demanding and do not have a direct interpretation, hindering user acceptance. To address these limitations, we introduce an approach for Personalised Recommendation of Visual arts based on learning latent semantic representation of paintings. Specifically, we trained a Latent Dirichlet Allocation (LDA) model on textual descriptions of paintings.  Our LDA model manages to successfully uncover non-obvious semantic relationships between paintings whilst being able to offer explainable recommendations. Experimental evaluations demonstrate that our method tends to perform better than exploiting visual features extracted using pre-trained Deep Neural Networks.

\end{abstract}

\IEEEpeerreviewmaketitle

\section{Introduction}
Paintings are important pieces in visual art that bring together complex elements such as drawings, gestures, narration, composition, abstraction, etc~\cite{mayer1991artist}. These elements carry deeper semantics beyond their usual categorizations (i.e. time period, material, size, color, etc.)  Paintings are also perceived differently by people as they trigger different emotional and cognitive reflections depending on the background and personality of a person~\cite{mayer1991artist}. Personalised recommendation of visual arts often suggest similar contents to those users have already seen or previously indicated that they liked. On the contrary, a recent work proposed by~\cite{frostart} presents an  Anti-Recommendation approach called \textit{``Art I don't like"}. This anti-recommender exposes users to a variety of content and suggests artworks that are dissimilar to the ones users selected aiming to allow serendipity and exploration. In comparison to other areas such as movie and music recommendation, Visual art has received little attention \cite{messina2017exploring}. Nevertheless, the huge potential and benefit of personalised recommendations of visual arts has been seen in the works of~\cite{messina2017exploring},~\cite{yilma2018introduction},~\cite{he2016vista} and~\cite{naudet2018personalisation}.
Similarities and relationships among paintings can normally be inferred based on common high level features such as colour, material, style, artist, etc. However, such features are not expressive enough as they cannot fully capture abstract concepts that are hidden in the paintings. In order to capture non-obvious contexts by a machine, effective data representation is very crucial \cite{bengio2013representation}. In general prominent works in visual art recommendation rely on ratings and  manually curated metadata (i.e. color, style, mood, etc.,) in order to drive recommendations. Recently works such as \cite{he2016vista} started to use latent visual features extracted using Deep Neural Networks (DNN) and also use pre-trained models for making visual art recommendations. According to results reported by~\cite{messina2017exploring}, DNN-based visual features perform better than manually curated metadata. However, these latent visual features do not have a direct interpretation and cannot be used to explain recommendations~\cite{verbert2013visualizing}. Hence they hinder user acceptance.

In this work we propose a personalised VA recommendation strategy by learning latent semantic representations. In particular we adopt a Latent Dirichlet Allocation (LDA) based representation learning approach \cite{blei2003latent.}. LDA is a topic analysis model that is known to be successful in the domain of Natural Language Processing (NLP) for uncovering hidden semantic relationships among documents.  In this approach paintings are represented by documents containing their detailed textual descriptions. Our LDA model is trained on 2368 paining collections from the National Gallery, London and employed for a personalised recommendation task.  The objective of this work is two-fold. Firstly, it introduces a strategy of leveraging textual data for visual art work recommendations. Secondly, it offers a method for driving explainable recommendations since the latent topics uncovered though LDA are directly interpretable. The rest of this paper is organised as follows. Section II covers a brief literature review of related works. Section III presents the problem formulation. Section IV illustrates the proposed solution introducing the LDA training and recommendation strategy. Section V presents an experimental evaluation of the proposed approach and a comparative analysis against visual feature based recommendations extracted using Deep Neural Networks. Finally Section V presents a concluding discussion.

\section{Related Work}
The earliest works in Information Retrieval (IR) and Natural language processing (NLP) have been using vector space models to represent documents as a vector of key words \cite{bengio2013representation}. However, such representations offered very limited reduction of description length and had a limited ability to capture inter/intra-document structures. To this end further techniques have been developed to tackle the curse of dimensionality by capturing hidden semantic structures in document modeling. In 2003 an unsupervised generative probabilistic model called Latent Dirichlet Allocation (LDA) was proposed~\cite{blei2003latent.}. LDA demonstrated superiority over the other models used at that time. Following this Latent variable models became widely accepted strategies to make inference about hidden semantic relationship between variables. Particularly in the domain of Recommendation Systems, LDA has been applied on several tasks such as online courses recommendation~\cite{apaza2014online}, personalized hashtag recommendation~\cite{zhao2016personalized}, scientific paper recommendation~\cite{amami2016lda}, similar TV user grouping and TV program recommendation~\cite{pyo2014lda}. LDA has been proven to be successful over several recommendation tasks. Hence, in this work we propose a strategy to leverage LDA in the domain of Visual Art recommendation.

\section{Problem formulation}
Let  $P=\{p_1, p_2, . . ., p_m\}$ be the set of paintings in a museum, $D=\{d_1, d_2, . . ., d_m\}$ be the set of documents containing textual descriptions of the paintings in $P$ and $P^u  = \{p^u_1, p^u_2, . . ., p^u_n\}$ be the set of paintings a user \textit{u} has rated where $P^u\subset P$. $W^u=\{w^u_1, w^u_2, . . ., w^u_n\}$ are the weights representing the rating given by user \textit{u} to the paintings in $P^u$. The task of personalised painting recommendation is to recommend a ranked list of paintings based on what paintings the user have liked or rated. However, considering the realistic scenario in a museum, access to user preferences is usually very limited.
Hence, providing a personalised recommendation service in such settings needs to rely on content-based approaches that can work well under limited preference information. In this work we propose a Latent Dirichlet Allocation (LDA) model, to build a content based personalised painting recommendation system. Our trained LDA model is capable of uncovering hidden semantic structures in the paintings to derive explainable recommendations. The LDA training and recommendation strategy is discussed in the next section. 
\section{Proposed Approach}
\begin{figure}[t]
\centering
\includegraphics[width=2.0in]{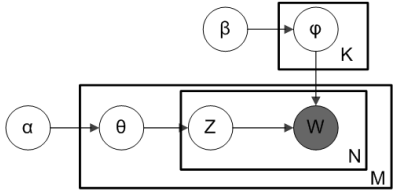}
\caption{ Latent Dirichlet Allocation plate diagram}
\label{fig:LDA_plate}
\end{figure}

\subsection{LDA Model}
Latent Dirichlet allocation (LDA) is an unsupervised generative probabilistic model for collections of text corpora proposed by \cite{blei2003latent.}.  The intuition behind LDA is that documents are represented as random mixtures over latent topics, where each topic is characterized by a distribution over words.  Our painting recommender system is built by creating a document description for each painting in the collection and training an LDA model on the corpus that can uncover semantic structures hidden in the paintings. This refers to everything that is not known a priori and hidden in the paintings.   
 Then the trained LDA model can be queried to recommend the most similar paintings for paintings a user have liked.

The intuition used in LDA is that each document can be seen as a combination of multiple topics. If we take paintings as an example, they can be described as a mixture of several concepts such as religion, nudity, portrait, etc. In LDA, each document is characterized by a predefined set of latent topics. In essence, each document is a distribution of topics and each topic is a distribution of words. This means each word in each document comes from a topic and the topic is selected from a per-document distribution over topics. Prominent words in each latent topic explain the nature of the topic and prominent latent topics related to each document explain the nature of the document (i.e. paintings). For instance, let us assume that latent topics are ``religion", ``still life", and ``landscape". A painting may have the following distribution over the topics : 70\%  ``religion", 10\% ``still life" and 20\% ``landscape". Moreover, each topic has a distribution over the words in the vocabulary. For topic ``religion", the probability of the word ``Saint" would be higher than in the topic ``landscape".
\begin{figure}[t]
\begin{centering}
\includegraphics[width=2.3in]{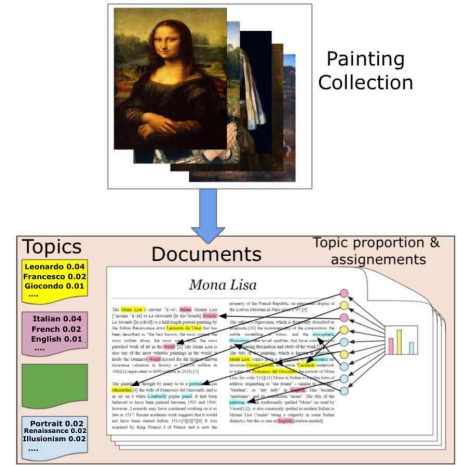}
\caption{ The intuition behind the Painting LDA model}
\label{fig:LDA_int}
\end{centering}
\end{figure}
The LDA model is represented as a probabilistic graphical model in Figure~\ref{fig:LDA_plate}.  $\alpha$ is the per-document  topic distribution, $\beta$ is the per-topic word distribution, $\theta$ is the topic distribution for a document $d$, $\varphi$ is the word distribution for the topic $K$ and $z$ the topic for the $n^{th}$ word in the document $d$ and finally, $W$ is a word. In LDA, each topic is a Multinomial distribution over the vocabulary in the collection of documents. To represent a topic, only the top-$n$ words are considered based on their probability. 
The procedure of building LDA model for a collection of paintings is explained as follows:
\begin{itemize}
    \item Pre-processing: we construct a collection of documents that contain detailed textual description of the paintings.
    \item Initialization:
    \begin{enumerate}
        \item We assume there is a defined number of topics $k$ in the collection of documents
        \item Attribute a topic to each word $W$ in the collection of documents where $\theta_{i} \sim Dir(\alpha)$ with $i \in \{1, ..., M\}$ and $Dir(\alpha)$ is a Dirichlet distribution. This initializes a topic model.
    \end{enumerate}
    \item Learning:
    \begin{enumerate}
\item We assume that the topic assigned to a word $W$ is wrong but that all the others are correct which consists in computing the conditional probabilities $p(topic\ t\ |\ document\ d)$ (probability that the document $d$ is assigned to the topic $t$) and $p(word\ w\ |\ topic\ t)$ (probability that the topic $t$ is assigned to the word $w$)
\item We update the topic of the document which is now the topic that has the highest probability to be assigned to this document ($p(topic\ t\ |\ document\ d) \cdot p(word\ w\ |\ topic\ t)$)
    \end{enumerate}
\end{itemize}

\begin{figure}[t]
\begin{centering}
\includegraphics[width=2.4in]{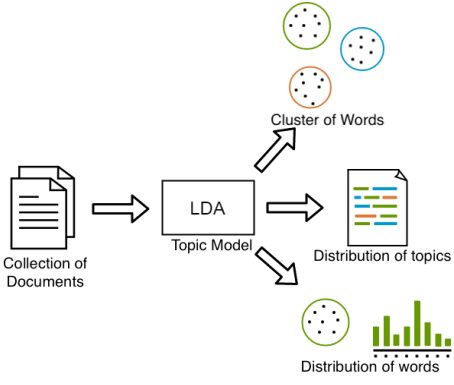}
\caption{ Topic modeling with LDA}
\label{fig:LDA_top}
\end{centering}
\end{figure}

Assigning the right number of topics as well as the hyperparameters however is not a trivial task. In literature the Gibbs sampling algorithm is widely used to estimate parameters of LDA \cite{ganguly2017paper2vec}\cite{grover2016node2vec}\cite{ng2017dna2vec}. In our implementation we used an (optimized version of) collapsed gibbs sampling from MALLET \cite{graham2012getting}. We refer the reader to the detailed discussion about LDA formulation in \cite{blei2003latent.} and \cite{jelodar2019latent}. Figure~\ref{fig:LDA_int} illustrates the intuition behind our Painting LDA model and  Figure~\ref{fig:LDA_top} explains topic modeling with LDA. As it is illustrated in the figures a collection of documents is used as an input to the LDA algorithm. LDA creates $k$ topics that can be seen as clusters of words. Each document of the collection is represented as a distribution of the topics which are themselves, distributions of words.

\subsection{Personalised Painting Recommendation }
\begin{figure}[t]
\begin{centering}
\includegraphics[width=2.4in]{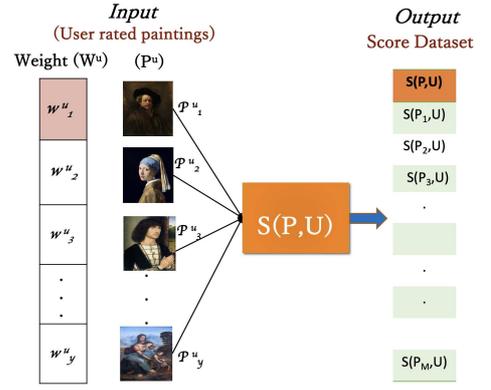}
\caption{ Painting Scoring Strategy}
\label{fig:LDA_Score}
\end{centering}
\end{figure}

Once the LDA model is trained over the entire corpus we get a matrix of documents by topics which expresses latent topic distribution of each painting. From this we can generate an $M$x$M$ similarity matrix for all the paintings in the dataset. 
For recommending paintings to a user $u$, user preferences towards paintings are modeled by a normalized weight vector, transformed from a 5 point Likert scale ratings: we assign a weight $w^u_i \in [0,1]$ for every painting $p^u_i$ a user has rated. The recommendation task is then to recommend most similar paintings to a user based on the set he has rated before. This is done by expanding user preferences towards unseen paintings and predicting a similarity score for the paintings in the dataset. The predicted score $S(p_i,u)$ for a painting $p_i$ in the dataset, according to the preferences of user $u$ is calculated based on the weighted average similarity score (distance) from all other paintings that have been rated by the active user given by:
\begin{equation}
 S(p_i,u) =  \frac{1}{N} \sum ^{N}_{j = 1} w_j * d(p_i,p_j) 
\end{equation}
where $d(p_i,p_j)$ is the similarity between painting $p_i$ and $p_j \in P^u$ according to the LDA similarity matrix. The summation is taken over all user rated paintings. This scoring strategy is illustrated on Figure~\ref{fig:LDA_Score}. Once the scoring procedure is complete, the paintings are sorted in a descending order and the first $K$ paintings are selected to generate a recommendation list. Our LDA based personalised painting recommendation procedure is sketched in Algorithm 1. 
\begin{algorithm}[H]
\caption{The {\sc LDA-based Personalised Painting recommender} algorithm; $\calR$ = \{\textit{$\vp_1$,... $\vp_k$}\} is a set of top k recommendations.} 
\begin{algorithmic}[1]
    \Procedure{LDA-based persoanlised Painting recommender}{$D,P, P^u, W^u$\calR$,S(p,u)$}
        \State ; Initialization
        \State ; repeat
        \For{$\vd_1$, \ldots, $\vd_m$}
            \State Train LDA model
                \For{$\vp_1, \ldots, \vp_m$}
                    \State Compute Cosine distance
        \EndFor
        \EndFor
        \State $\calR \gets \emptyset$
        \State ; repeat 
        \For{$\vp_1$, \ldots, $\vp_m$}
            \State $S(p_i,u)$ $\gets$ $\Call{Compute}{\frac{1}{N} \sum ^{N}_{j = 1} w_j * d(p_i,p_j)}$
        \EndFor
            
            \State Sort $ $\textit{P}$ \gets \Call{Descending}{S(p_i,u)}$
        
        \State \Return $\{\textit{$\vp_1$,... $\vp_k$}\} \gets \Call{SELECT Top k}{\calR} $
    \EndProcedure
\end{algorithmic}
\end{algorithm}

\section{Experiment \& Discussion}
This section presents our experimental results. First, we briefly describe the dataset. Then, we introduce the metric used to evaluate the proposed topic model, and present an analysis regarding the capability of our approach in terms of explainability. Finally, we describe our method to assess the quality of our approach in the context of a recommendation task.

\subsection{Dataset}
In this study we used a dataset containing 2368 paintings from the National Galary of London. Each painting is represented by a set of attributes which are summarized in Table I. For our approach we specifically focus on painting description attribute. These descriptions are provided by museum experts and carry complementary information about the paintings such as stories and narratives that can be exploited to capture the semantic of a painting. These descriptions provide concise information about each painting. For the task of topic modeling using LDA, we decided to enrich the descriptions dataset $D$ by concatenating the paintings descriptions with keyword attributes such as the artist name, the painting title, the technique used, the publication date, the format (landscape, portrait) and the size (small, medium, etc.,) and also additional information from other sources to better train the model. Hence we generated a second dataset $DE$ enriched with additional textual descriptions and stories related to the paintings, collected from various sources such as Wikipedia and books. To avoid ``Garbage in, garbage out" we performed textual pre-processing on both datasets (i.e. removal of punctuation, stop words, bi-grams, Lemmatization) as they do not add any value to the topic models.

\begin{table}[h]
\parbox{.99\linewidth}{
\centering
\begin{tabular}{|l|1|}
\hline
\textbf{Attributes} & \textbf{Description}\\
\hline
artist\_name & Artist name\\
\hline
painting\_title & Painting title\\
\hline
painting\_id & Painting identifier\\
\hline
painting\_description & Description \\
\hline
publishing\_date & Publication date\\
\hline
format & (Landscape, Portrait)\\
\hline
size & (Small, Medium, Large)\\
\hline
technique & (Oil, tempera, ...)\\
\hline
\end{tabular}
}
\\
\caption{Attributes \& descriptions of the Paintings dataset}
\end{table}
\subsection{Model Evaluation}
\begin{figure}[t]
\begin{centering}
\includegraphics[width=2.7in]{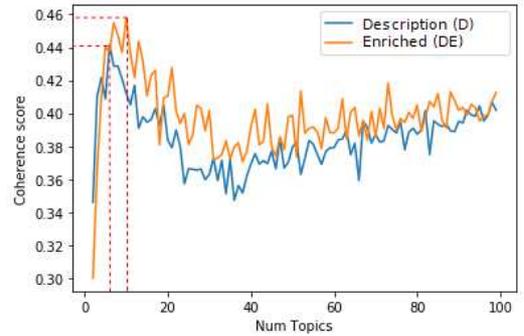}
\caption{ Comparative topic coherence analysis of the two data sets}
\label{fig:Top_COH}
\end{centering}
\end{figure}

In topic modeling, \textit{Topic Coherence} is a commonly used technique to evaluate topic models. It is defined as the sum of pairwise similarity scores on the words $w_1, ..., w_n$ used to describe the topic, usually the top $n$ words by frequency $p(w|t)$ \cite{jelodar2019latent,newman2010automatic}.
\begin{equation}
 Coherence = \sum ^{}_{i < j} Score (w_i,w_j) 
\end{equation}
Ideally, a good model should generate coherent topics. (i.e the higher the coherence score the better the topic model is \cite{newman2010automatic}). In order to identify the optimal number of topics as well as which data set $D$ or $DE$ generates the best topic model, our implementation resorted the topic coherence pipeline from gensim library\footnote{https://radimrehurek.com/gensim/models/ldamodel.html} \textit{CoherenceModel}. Figure \ref{fig:Top_COH} shows the evolution of the topic coherence as a function of the number of topics for each of the two dataset. 
From the analysis presented in figure \ref{fig:Top_COH} we can make two observations. Firstly, the data set $DE$ (orange) generally gives slightly better topic coherence score and thus, a better topic model. Secondly, we can observe that with 10 topics we obtained a topic coherence of approximately 0.46 for the data set $DE$ which is the best score for this window. Having too many topics requires more resources as well as time for a result that is not significantly better. Thus, we decided to limit the number of topics to 10. Generally the topic coherence scores shows that enriching the original descriptions with additional information led to a better topic model. Hence, we chose to work with the dataset $DE$ instead of $D$. 

\begin{figure}[t]
\begin{centering}
\includegraphics[width= 1.5in]{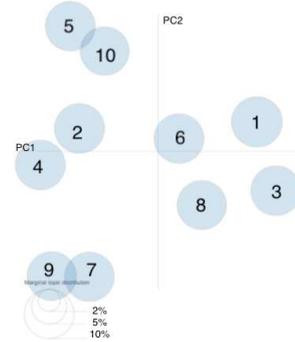}
\caption{Inter-topic Distance Map in a 2-dimensional space of 10 topics}
\label{fig:Vis}
\end{centering}
\end{figure}

In addition to the evaluation of the topic coherence, we visualize the topics by using a visualization tool called pyLDAvis\footnote{https://pypi.org/project/pyLDAvis/}. In figure \ref{fig:Vis}, we see each topic represented by a circle annotated with a digit from 0 to 9. The size of the circle represents the prevalence of a topic, i.e, the popularity of a topic among the paintings. The distance represents the similarity between topics. In fact it is an approximation to the original topic similarity because we are using a 2-dimensional scatter plot to best represent the spatial distribution of the topics. The objective here is to have topics that are overlapping as little as possible. With 10 topics, we can observe that the topics are evenly popular while being sufficiently distinct from each other. This means that the topics are sufficiently different which is what it is intended. 

\subsection{Explainability}
In figure 7.a and 7.b we present two paintings\footnote{These paintings are available under Creative Commons License}. The painting in 7.b) is the most similar painting to the painting 7.a) based on our LDA topic model. We observe in figure 7.c and 7.d that their topic distribution is very similar and one particular topic stands out: topic 8. The fact that this topic stands out from the others for these paintings implies that the words found in this topic are more likely to be found in the paintings descriptions than the words from the other topics. 
\begin{table*}[]
\centering
\label{table2}
\begin{tabular}{ccc}
\includegraphics[width= 3cm]{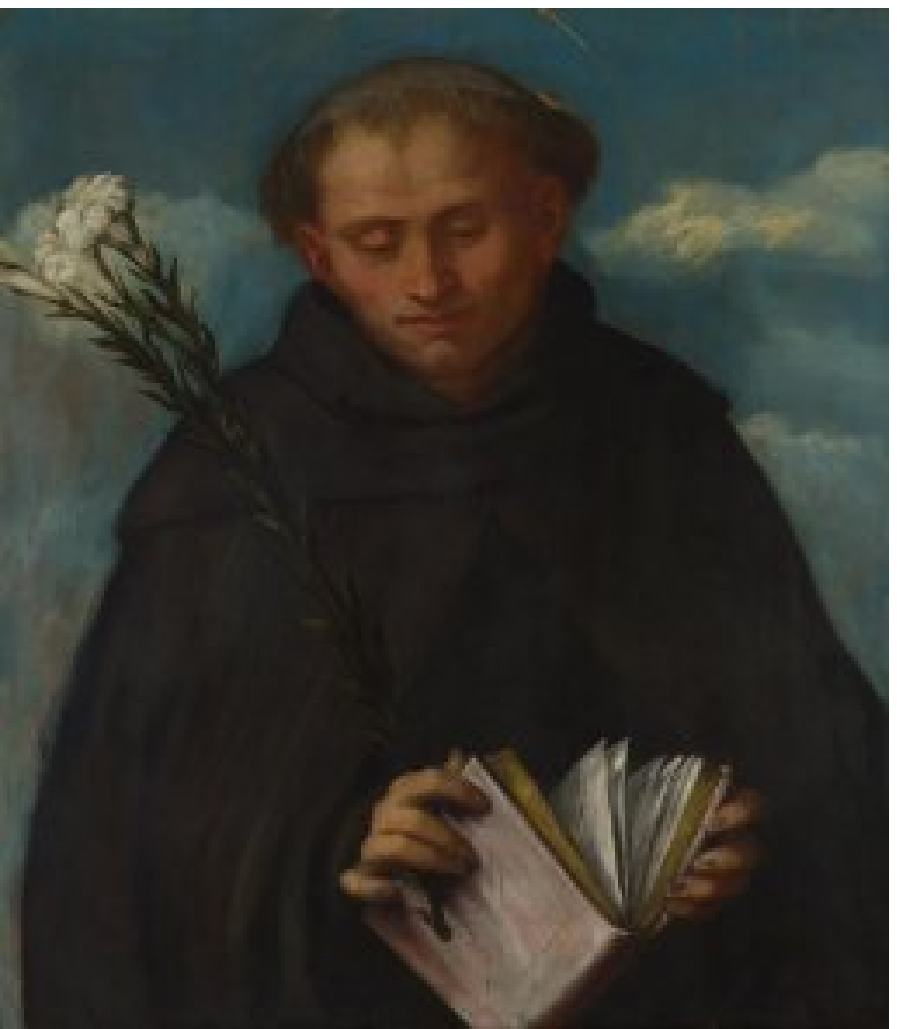}                  & \includegraphics[width= 3cm]{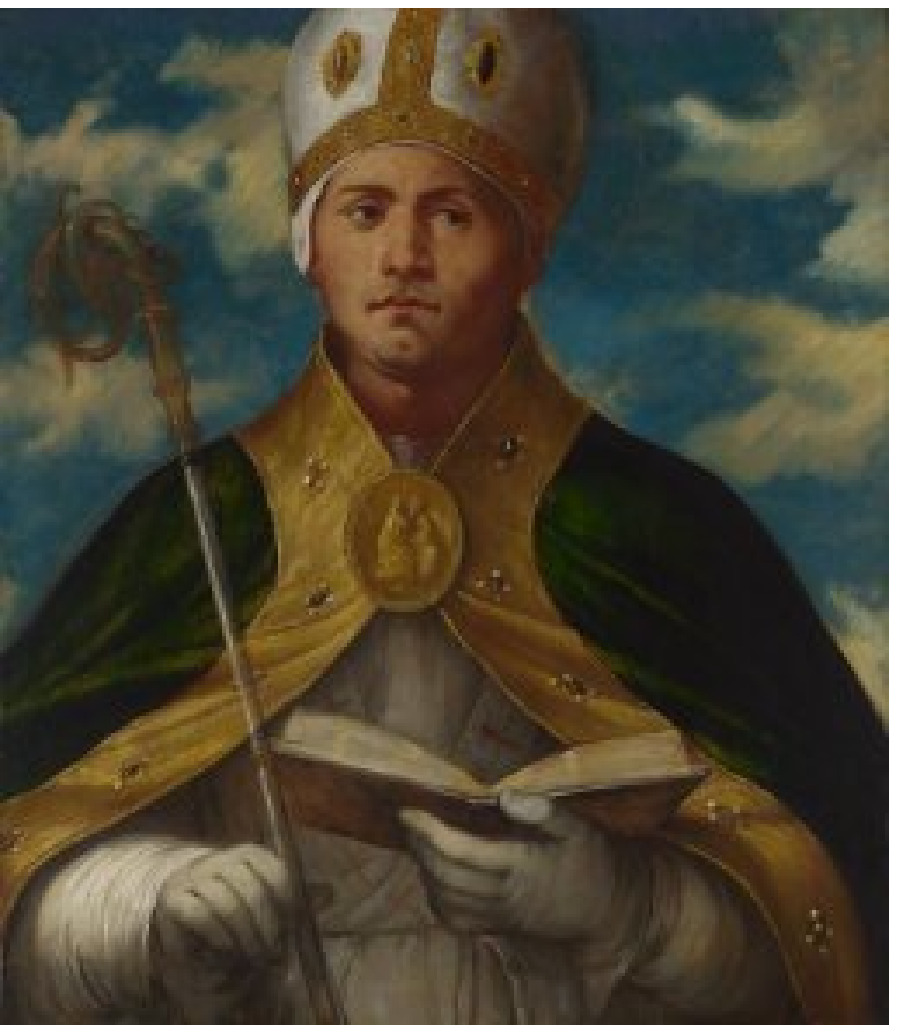}  
\\ 
(a) Target Painting & (b) Most similar Painting
\\
\includegraphics[width= 5.7cm]{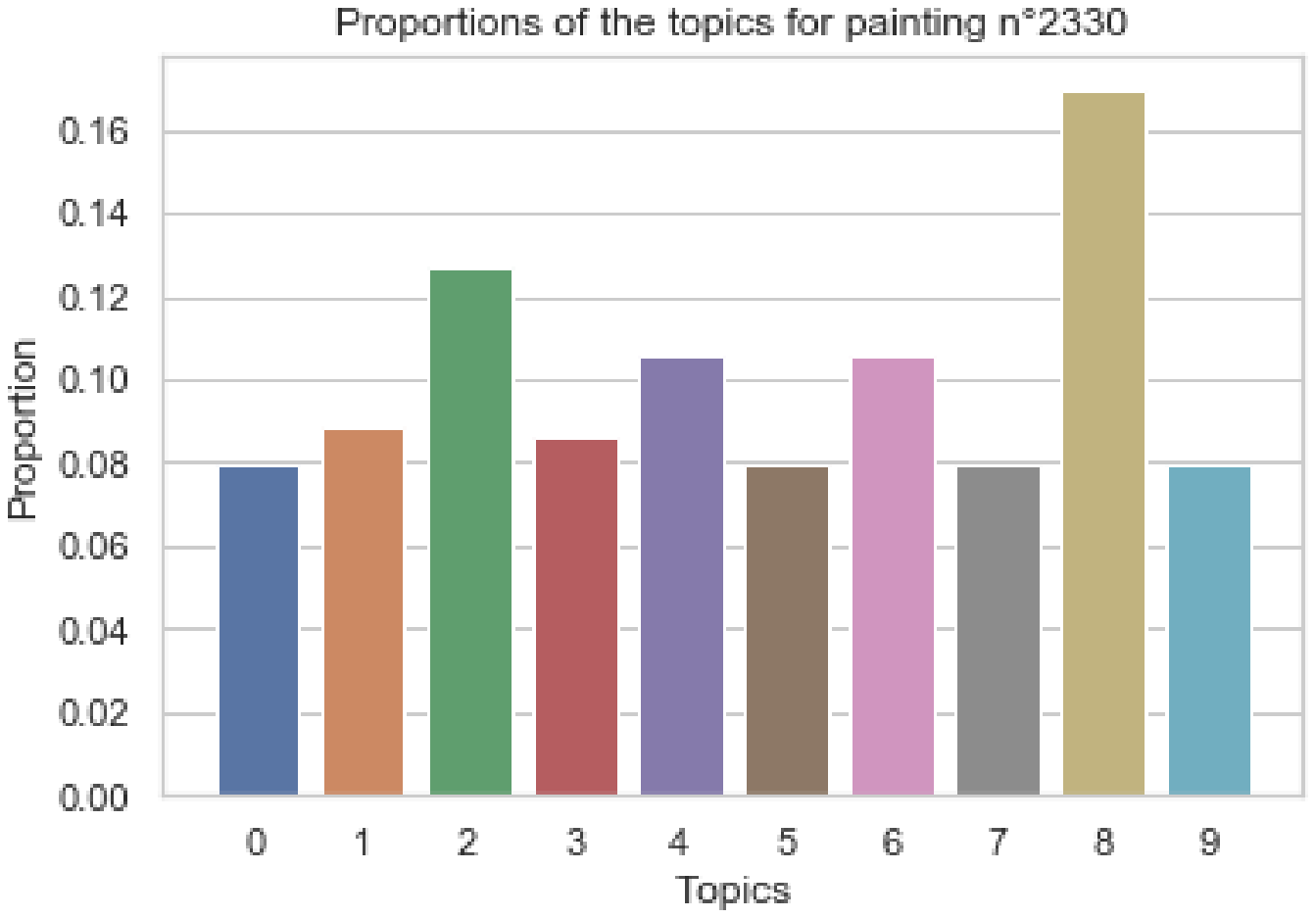}   & \includegraphics[width= 5.7cm]{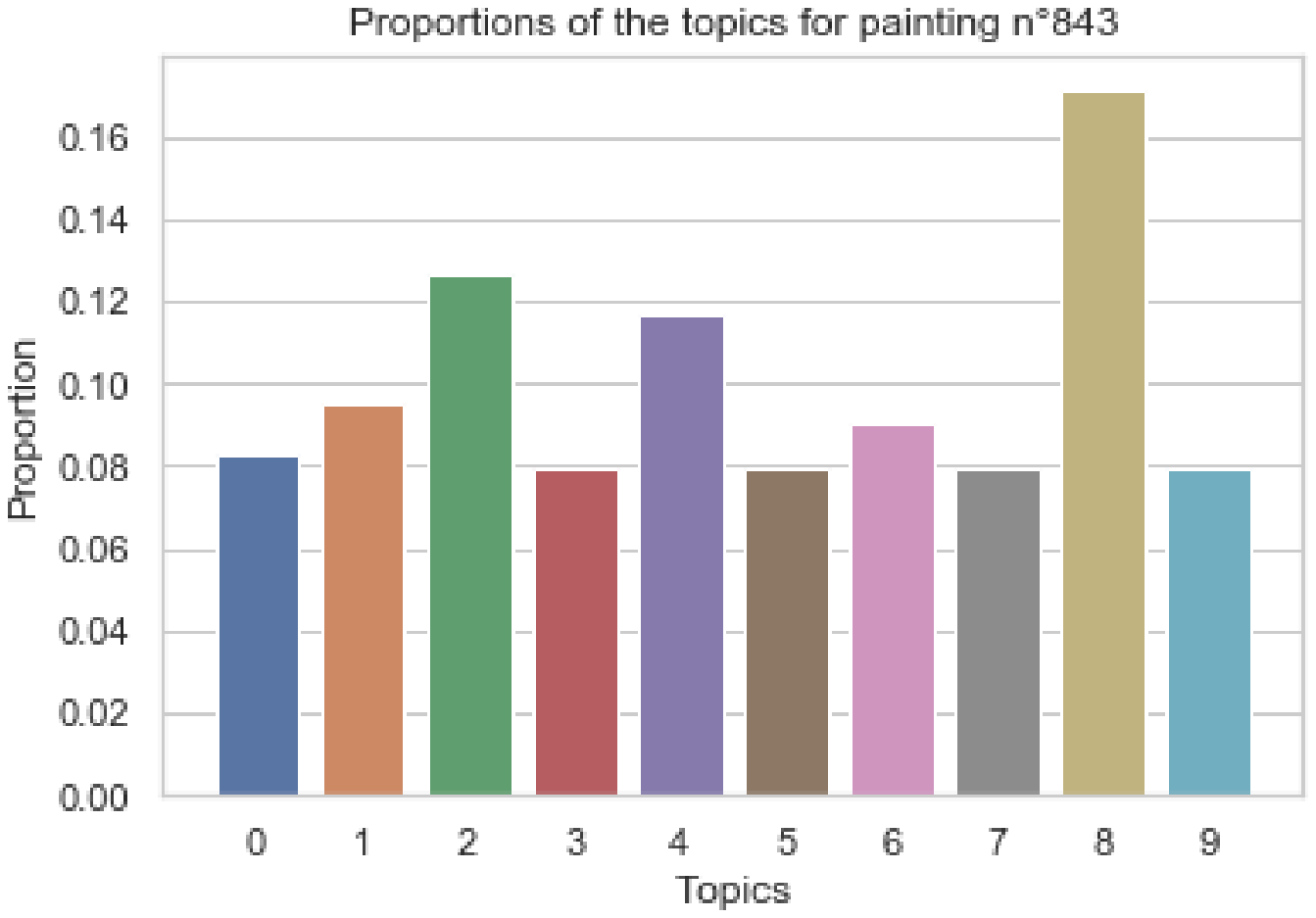} 
\\ 
(c) Topic distribution of target painting & (d) Topic distribution of the most similar painting
\end{tabular}
\begin{flushleft}

Fig.7.  Target painting (Top left), its most similar painting (Top right), Topic distribution of the target (bottom left) and its most similar painting (bottom right).
\end{flushleft} 
\end{table*} \\
When we have a closer look at the topics descriptions in Table II, we can see that the topic 8 is very well defined as it is obvious that there is a coherence between the words that are used. In fact, topic 8 can be described as the ``christian" topic since many of the words in this topic are usually found in christian corpora. When we look at the paintings, it is obvious that there are many references to Christianity and therefore, we can assume that their descriptions contain vocabulary that refers to a religious context. On the other hand, when we look at the descriptions of topics 0 and 5, what is interesting is that the words found in these topics do not seem to share the same semantic at first sight. However, the underlying intuition behind LDA is that words are in the same topics because they are \textit{frequently} found together and therefore, they are used in a same context. Thus, words ``Paris" and ``flower" found in the topic 5 that do not seem to have any obvious relation are nonetheless in the same topic because many paintings descriptions in our data set describe Paris and flowers together. LDA is able to find relations between words and at a broader extent, paintings that are not always obvious to us but explainable. Therefore for the task of recommendation using LDA we can easily identify prominent topics and reason out shared semantics among recommendations in order to help users better understand recommendations. In our experiment we offered explanations to users as a word cloud of prominent concepts (i.e. topics extracted with LDA) shared among the recommendations matching their personal preferences. Fig. 8 is an example of interest word cloud. 
\begin{figure}[t]
\centering
\includegraphics[width=2.0in]{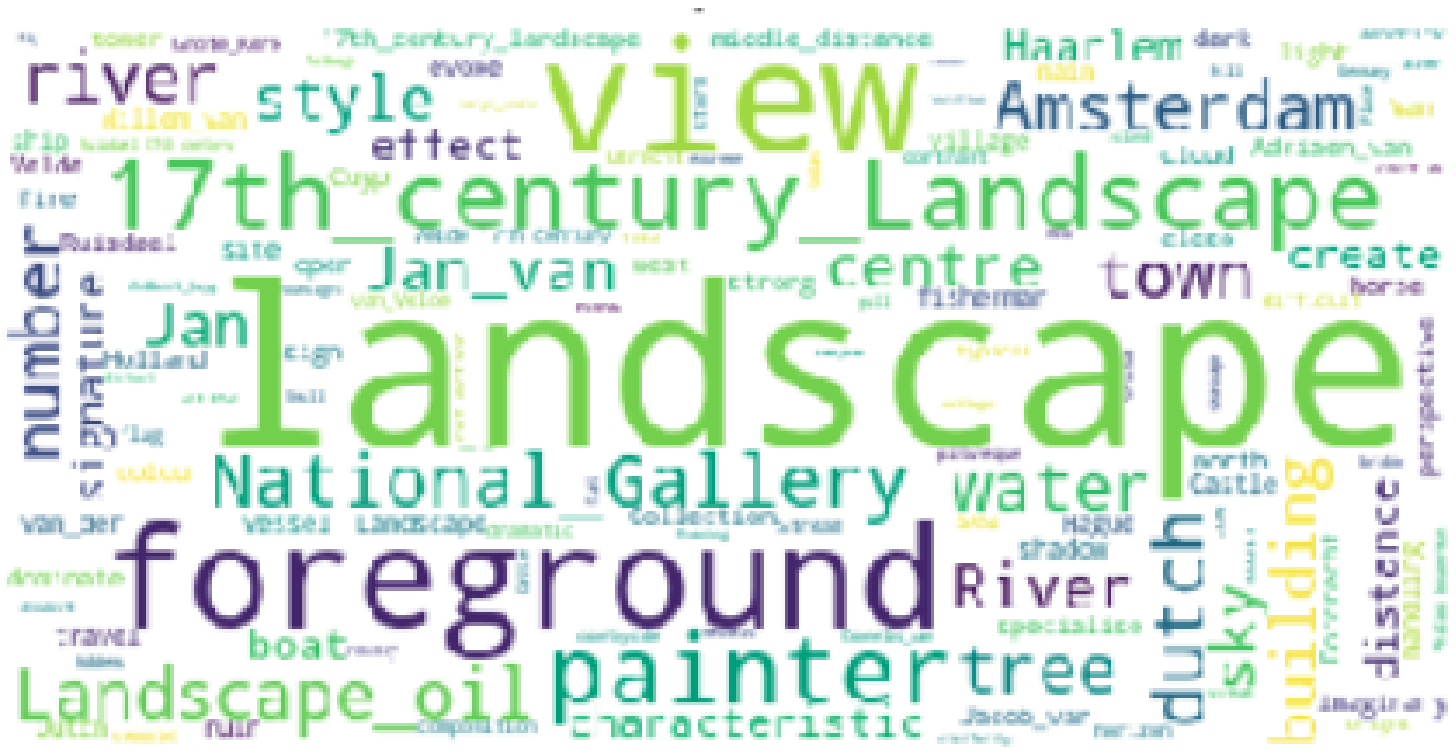}
\captionsetup{labelformat=empty}
\caption{Fig. 8. Example of prominent concepts shared among recommendations (i.e explanation)}
\label{fig:land}
\end{figure}

\begin{table}
\parbox{.99\linewidth}{
\centering
\begin{tabular}{|l|1|1|}
\hline
\textbf{Topic 0} & \textbf{Topic 5}&
\textbf{Topic 8}\\
\hline
LANDSCAPE & PARIS &  CHRIST \\
\hline
SCENE & LIGHT & SAINT\\
\hline
VIEW & FORM & ALTARPIECE\\
\hline
17th C LANDSCAPE &COLOUR & PANEL\\
\hline
PEASANT & SKETCH & JESUS\\
\hline
DUTCH & STUDIO & NEW TESTAMENT\\
\hline
LANDSCAPE OIL & FLOWER & EVANGELIST\\
\hline
TREE & COMPOSITION & CROSS\\
\hline
TOWN & STUDY & CHURCH\\
\hline
RIVER & 19th CENTURY LANDSCAPE & CRUCIFICATION\\
\hline
\end{tabular}}
\\
\caption{Description of three topics}
\end{table}
\subsection{Recommendation Quality}
 In order to evaluate the performance of our algorithm on a recommendation task we performed a controlled experiment with 15 real users. Each user provided a rating for 80 paintings from the dataset to be used for profiling. As a baseline recommender we used visual features extracted using Residual Networks (ResNet-50) \cite{he2016deep}. ResNet-50  is a  50-layer deep convolutional neural network trained on more than a million images from the ImageNet database \footnote{ImageNet. http://www.image-net.org}. Thus, it has learned rich feature representations for a wide range of images. We used the pre-trained ResNet-50 to extract latent visual features from our painting dataset. The extracted features are then used to identify similar paintings to user preferred ones through similar scoring mechanism used for LDA. The \textit{User-Centric} evaluation was done through a questionnaire where participants had to express their opinion in a Five point Likert scale to four of the following statements. \textit{``The recommendations match my personal preferences and interests"} (Predictive accuracy); \textit{``The recommender helped me discover paintings I did not know before"} (Novelty); \textit{``The recommender helped me discover surprisingly interesting paintings I might not known otherwise."} (Serendipity); \textit{``The recommended paintings are diverse"} (Diversity). Additionally users were asked if explanations offered by LDA helped them to better understand recommendations. Interestingly all participants responded \textit{``Yes"}, this shows the tendency that explainable recommendations have a positive impact on user experience. Figure 9 summarizes the average values of user ratings for the two recommendation pipelines.  As reported on the figure, LDA achieved significantly higher diversity values (4/5) compared to ResNet-50 (2.3/5). This is due to the fact that the notion of similarity in LDA is directly related to semantically dominant concepts shared among paintings. i.e. LDA also uncovers relationship between paintings that do not necessarily have a resembling visual features. Hence, LDA captures semantic similarities that cannot be discovered through latent visual features. This can also justify the higher values of serendipity and Novelty since LDA based recommendation can contain visually diverse but semantically related paintings. In terms of matching user preferences, LDA also performs significantly better (4.1/5) than ResNet-50 (3.1/5). The preliminary results presented in this section demonstrate the potential of our algorithm to benefit the Visual Art recommendation domain. Hence, moving forward further experimental evaluation on a larger and diverse set of users should be conducted. 
\begin{figure}[t]
\centering
\includegraphics[width=3.0in]{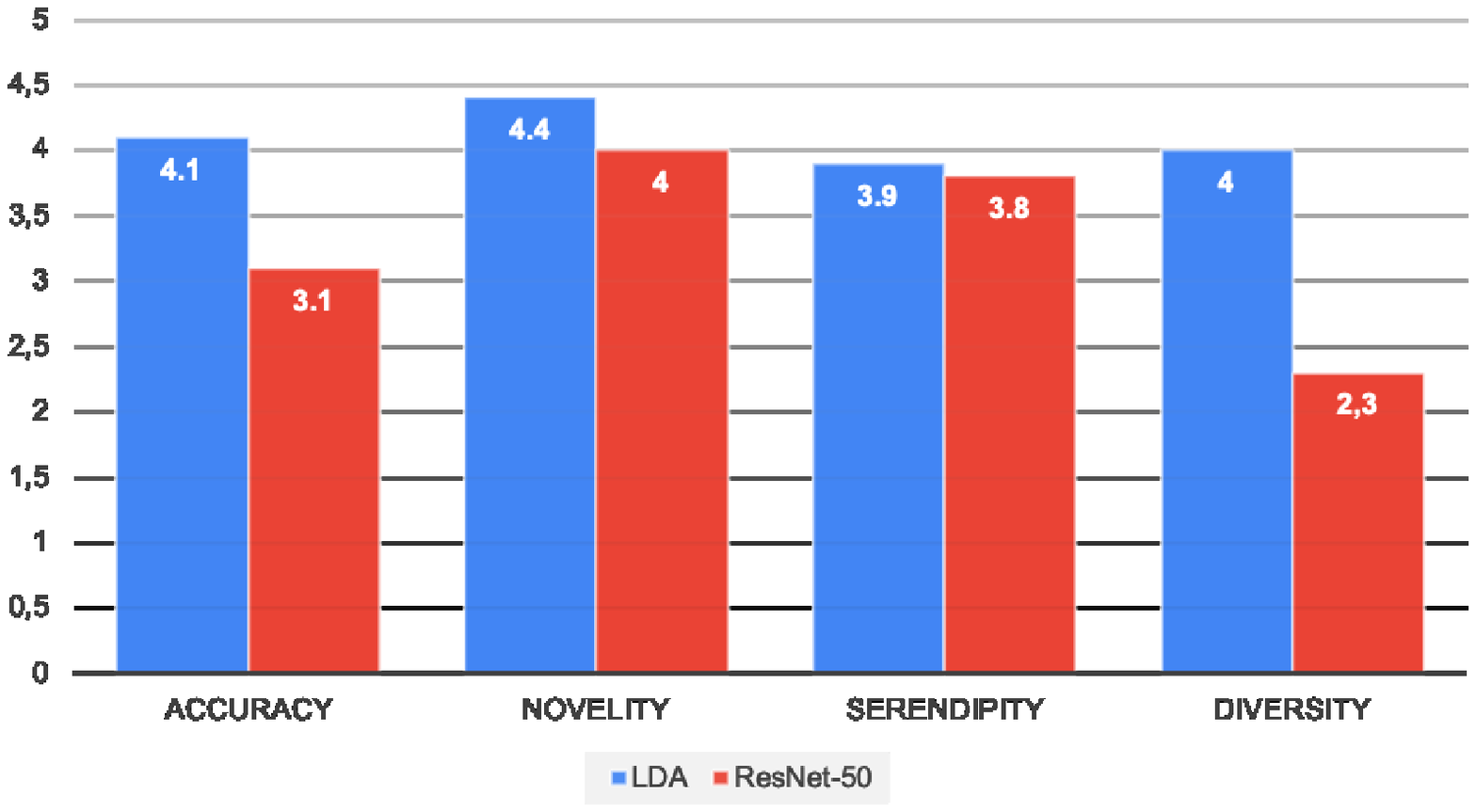}
\captionsetup{labelformat=empty}
\caption{Fig. 9. Comparison of LDA Vs ResNet-50 interms of \textit{Accuracy, Novelity,	Serendipity	and Diversity}}
\label{fig:ucr}
\end{figure}
\section{Conclusion}
In this work we proposed a personalised Visual Art recommendation strategy based on learning latent semantic representations. In particular we trained an LDA based algorithm which leverages textual descriptions of paintings to uncover latent semantic relationships among paintings. Our algorithm compares favourably with state of the art VA recommendation method based on visual features extracted using Deep Neural Networks. The latent features extracted with LDA can directly be interpreted by humans. Thus, it enables a personaliser to offer explainable recommendation which is proved to be helpful for users to better understand recommendations.
With 15 users sample, our approach shows better results than the DNN approach. Interestingly, this is true for the 4 dimensions of recommendations we have tested (accuracy, novelty, serendipity, diversity). This result needs however to be further confirmed, as well as the tentative explanations we have given. As a future work, further validation with larger set of user groups and also a hybrid approach combining visual features could be a relevant extension.

\bibliographystyle{IEEEtran}
\bibliography{IEEEabrv,SMAP2020}

\begin{thebibliography}{10}
\providecommand{\url}[1]{#1}
\csname url@samestyle\endcsname
\providecommand{\newblock}{\relax}
\providecommand{\bibinfo}[2]{#2}
\providecommand{\BIBentrySTDinterwordspacing}{\spaceskip=0pt\relax}
\providecommand{\BIBentryALTinterwordstretchfactor}{4}
\providecommand{\BIBentryALTinterwordspacing}{\spaceskip=\fontdimen2\font plus
\BIBentryALTinterwordstretchfactor\fontdimen3\font minus
  \fontdimen4\font\relax}
\providecommand{\BIBforeignlanguage}[2]{{%
\expandafter\ifx\csname l@#1\endcsname\relax
\typeout{** WARNING: IEEEtran.bst: No hyphenation pattern has been}%
\typeout{** loaded for the language `#1'. Using the pattern for}%
\typeout{** the default language instead.}%
\else
\language=\csname l@#1\endcsname
\fi
#2}}
\providecommand{\BIBdecl}{\relax}
\BIBdecl

\bibitem{mayer1991artist}
R.~Mayer, \emph{The artist's handbook of materials and techniques}, 1991.

\bibitem{frostart}
\BIBentryALTinterwordspacing
S.~Frost, M.~M. Thomas, and A.~G. Forbes, ``Art i don’t like: An
  anti-recommender system for visual art,'' 2019. [Online]. Available:
  \url{https://mw19.mwconf.org/paper/art-i-dont-like-an-anti-recommender-system-for-visual-art/}
\BIBentrySTDinterwordspacing

\bibitem{messina2017exploring}
P.~Messina, V.~Dominguez, D.~Parra, C.~Trattner, and A.~Soto, ``Exploring
  content-based artwork recommendation with metadata and visual features,''
  \emph{arXiv preprint arXiv:1706.05786}, 2017.

\bibitem{yilma2018introduction}
B.~A. Yilma, Y.~Naudet, and H.~Panetto, ``Introduction to personalisation in
  cyber-physical-social systems,'' in \emph{OTM Confederated International
  Conferences" On the Move to Meaningful Internet Systems"}.\hskip 1em plus
  0.5em minus 0.4em\relax Springer, 2018, pp. 25--35.

\bibitem{he2016vista}
R.~He, C.~Fang, Z.~Wang, and J.~McAuley, ``Vista: a visually, socially, and
  temporally-aware model for artistic recommendation,'' in \emph{Proceedings of
  the 10th ACM Conference on Recommender Systems}, 2016, pp. 309--316.

\bibitem{naudet2018personalisation}
Y.~Naudet, B.~A. Yilma, and H.~Panetto, ``Personalisation in cyber physical and
  social systems: the case of recommendations in cultural heritage spaces,'' in
  \emph{2018 13th International Workshop on Semantic and Social Media
  Adaptation and Personalization (SMAP)}.\hskip 1em plus 0.5em minus
  0.4em\relax IEEE, 2018, pp. 75--79.

\bibitem{bengio2013representation}
Y.~Bengio, A.~Courville, and P.~Vincent, ``Representation learning: A review
  and new perspectives,'' \emph{IEEE transactions on pattern analysis and
  machine intelligence}, vol.~35, no.~8, pp. 1798--1828, 2013.

\bibitem{verbert2013visualizing}
K.~Verbert, D.~Parra, P.~Brusilovsky, and E.~Duval, ``Visualizing
  recommendations to support exploration, transparency and controllability,''
  in \emph{Proceedings of the 2013 international conference on Intelligent user
  interfaces}, 2013, pp. 351--362.

\bibitem{blei2003latent.}
D.~M. Blei, A.~Y. Ng, and M.~I. Jordan, ``Latent dirichlet allocation,''
  \emph{Journal of machine Learning research}, vol.~3, no. Jan, pp. 993--1022,
  2003.

\bibitem{apaza2014online}
R.~G. Apaza, E.~V. Cervantes, L.~C. Quispe, and J.~O. Luna, ``Online courses
  recommendation based on lda.'' in \emph{SIMBig}.\hskip 1em plus 0.5em minus
  0.4em\relax Citeseer, 2014, pp. 42--48.

\bibitem{zhao2016personalized}
F.~Zhao, Y.~Zhu, H.~Jin, and L.~T. Yang, ``A personalized hashtag
  recommendation approach using lda-based topic model in microblog
  environment,'' \emph{Future Generation Computer Systems}, vol.~65, pp.
  196--206, 2016.

\bibitem{amami2016lda}
M.~Amami, G.~Pasi, F.~Stella, and R.~Faiz, ``An lda-based approach to
  scientific paper recommendation,'' in \emph{International conference on
  applications of natural language to information systems}.\hskip 1em plus
  0.5em minus 0.4em\relax Springer, 2016, pp. 200--210.

\bibitem{pyo2014lda}
S.~Pyo, E.~Kim \emph{et~al.}, ``Lda-based unified topic modeling for similar tv
  user grouping and tv program recommendation,'' \emph{IEEE transactions on
  cybernetics}, vol.~45, no.~8, pp. 1476--1490, 2014.

\bibitem{ganguly2017paper2vec}
S.~Ganguly and V.~Pudi, ``Paper2vec: Combining graph and text information for
  scientific paper representation,'' in \emph{European Conference on
  Information Retrieval}.\hskip 1em plus 0.5em minus 0.4em\relax Springer,
  2017, pp. 383--395.

\bibitem{grover2016node2vec}
A.~Grover and J.~Leskovec, ``node2vec: Scalable feature learning for
  networks,'' in \emph{Proceedings of the 22nd ACM SIGKDD international
  conference on Knowledge discovery and data mining}, 2016, pp. 855--864.

\bibitem{ng2017dna2vec}
P.~Ng, ``dna2vec: Consistent vector representations of variable-length
  k-mers,'' \emph{arXiv preprint arXiv:1701.06279}, 2017.

\bibitem{graham2012getting}
S.~Graham, S.~Weingart, and I.~Milligan, ``Getting started with topic modeling
  and mallet,'' The Editorial Board of the Programming Historian, Tech. Rep.,
  2012.

\bibitem{jelodar2019latent}
H.~Jelodar, Y.~Wang, C.~Yuan, X.~Feng, X.~Jiang, Y.~Li, and L.~Zhao, ``Latent
  dirichlet allocation (lda) and topic modeling: models, applications, a
  survey,'' \emph{Multimedia Tools and Applications}, vol.~78, no.~11, pp.
  15\,169--15\,211, 2019.

\bibitem{newman2010automatic}
D.~Newman, J.~H. Lau, K.~Grieser, and T.~Baldwin, ``Automatic evaluation of
  topic coherence,'' in \emph{Human language technologies: The 2010 annual
  conference of the North American chapter of the association for computational
  linguistics}.\hskip 1em plus 0.5em minus 0.4em\relax Association for
  Computational Linguistics, 2010, pp. 100--108.

\bibitem{he2016deep}
K.~He, X.~Zhang, S.~Ren, and J.~Sun, ``Deep residual learning for image
  recognition,'' in \emph{Proceedings of the IEEE conference on computer vision
  and pattern recognition}, 2016, pp. 770--778.

\end{thebibliography}

\end{document}